\documentclass[showpacs,preprintnumbers]{revtex4-2}
\usepackage{amssymb}
\usepackage{graphicx}
\usepackage{color}
\usepackage{xcolor}

\def\bc{\begin{center}}
\def\ec{\end{center}}

\def\beq{\begin{equation}}
\def\eeq{\end{equation}}

\usepackage{graphicx}
\usepackage{caption}
\usepackage{subcaption}
\usepackage[T1]{fontenc}
\usepackage[utf8]{inputenc}
\usepackage[english]{babel}
\usepackage[hidelinks]{hyperref}
\usepackage{indentfirst}
\usepackage{amssymb}
\usepackage{amsmath}
\usepackage{mathrsfs}
\usepackage{braket}
\usepackage{float}
\usepackage{color}
\usepackage{setspace}

\begin{document}

\title{Superfluidity of dipolar excitons in a double layer of $\alpha-T_{3}$ with a mass term}

\author{Oleg L. Berman$^{1,2}$,   Godfrey Gumbs$^{2,3,4}$, Gabriel P. Martins$^{1,2,3}$, and Paula Fekete$^{5}$}
\affiliation{\mbox{$^{1}$Physics Department, New York City College
of Technology, The City University of New York,} \\
Brooklyn, NY 11201, USA \\
\mbox{$^{2}$The Graduate School and University Center, The
City University of New York,} \\
New York, NY 10016, USA\\
\mbox{$^{3}$Department of Physics and Astronomy, Hunter College of
The City University of New York
City University of New York,} \\
New York, NY 10065, USA\\
\mbox{$^{4}$Donastia International Physics Center (DIPC),
P de Manuel Lardizabal, 4, 20018 San Sebastian, Basque Country, Spain}\\
\mbox{$^{5}$ US Military Academy at West Point, 606 Thayer Road,
West Point, New York 10996, USA} }

\date{\today}

\begin{abstract}

We predict Bose-Einstein condensation and superfluidity of dipolar excitons,
formed by electron-hole pairs in spatially separated gapped hexagonal $\alpha-T_{3}$ (GHAT3) layers. In the  $\alpha-T_{3}$ model,
the AB-honeycomb lattice structure is supplemented with C atoms located at the centers of the hexagons in the lattice.
We considered the $\alpha-T_{3}$ model in the presence of a mass term which opens a gap in the energy dispersive spectrum.
The gap opening mass term, caused by a weak magnetic field, plays the role  of Zeeman splitting at low magnetic fields for this pseudospin-1 system.
The band structure of GHAT3 monolayers leads to the formation of two distinct types of excitons in the GHAT3 double layer.
We consider two types of dipolar excitons in double-layer GHAT3: (a) ``A excitons'', which are bound states of electrons
in the conduction band (CB) and holes in the intermediate band (IB) and (b) ``B excitons'', which are bound states of electrons in the CB and holes in the valence band (VB).
 The binding energy of A and B dipolar excitons is calculated. For a two-component weakly interacting Bose gas of
 dipolar excitons in a GHAT3 double layer, we obtain the energy dispersion of collective excitations, the sound velocity,
 the superfluid density, and the mean-field critical temperature $T_{c}$ for superfluidity.

\end{abstract}

\maketitle

\section{Introduction}
\label{intro}

The many-particle systems of dipolar (indirect) excitons, formed by spatially separated electrons and holes, in semiconductor coupled quantum wells (CQWs) and novel two-dimensional (2D) materials have been the subject of numerous experimental and theoretical studies. These systems are attractive in large part due to the possibility of Bose-Einstein condensation (BEC) and superfluidity of dipolar excitons, which can be observed  as persistent electrical currents in each quantum well, and also through coherent optical properties~\cite{Lozovik,Snoke,Butov,Eisenstein}. Recent progress in theoretical and experimental studies of BEC and superfluidity of dipolar excitons in CQWs have been reviewed in Ref.~\cite{Snoke_review}. Electron-hole superfluidity  in double layers can occur not only in the BEC regime, but also in the Bardeen-Cooper-Schrieffer (BCS)-BEC crossover regime~\cite{Hamilton}.

\medskip
\par

A number of experimental and theoretical investigations have been devoted to  the BEC of electron-hole pairs, formed by spatially separated electrons and holes  in a double layer formed by parallel graphene layers.   These investigations were reported in Refs.~\cite{BLG,Sokolik,Bist,BKZg,Perali}. Both BEC and superfluidity of dipolar excitons in double layers of transition-metal dichalcogenides (TMDCs)~\cite{Fogler,MacDonald_TMDC,BK,BK2,Conti} and phosphorene~\cite{BGK,Peeters} have been discussed, because the exciton binding energies in novel  2D semiconductors is quite large. Possible BEC in a long-lived dark spin state of 2D  dipolar excitons has been experimentally observed for GaAs/AlGaAs semiconductor CQWs~\cite{Rapaport}.

\medskip
\par
Recently, the electronic properties of the $\alpha-T_{3}$ lattice have been the subject of the intensive theoretical and experimental investigations due to its surprising fundamental physical properties as well as its promising applications in solid state devices~\cite{f1,f2,f3,f4,f5,f6.1,f6,f7,f8,f9,RKKY,t1,t2,t3}. For a review of artificial flat band systems, see Ref.\ \cite{Review}. Raoux, et al. \cite{f1}  proposed  that an $\alpha-T_{3}$  lattice could be assembled  from cold fermionic atoms  confined to an optical lattice by means of three pairs of laser beams for the optical dice lattice ($\alpha=1$) \cite{21}. This structure consists of an AB-honeycomb lattice (the rim)  like that in graphene which is combined with C atoms at the center/hub of each hexagon.  A parameter $\alpha$ represents the ratio of the hopping integral between the hub and the rim to that around the rim of the hexagonal lattice. By dephasing one of the three pairs of laser beams, one could vary the parameter $0 \leq \alpha =\tan\ \phi \leq 1$. Optically induced dressed states~\cite{AI1}, and their tunneling, transport~\cite{AI2,AI3}, and collective properties~\cite{AI5}, as well as $\alpha-T_{3}$ based nanoribbons~\cite{AI6} have been analyzed.  The BEC and superfluidity of dipolar magnetoexcitons in   $\alpha-T_{3}$ double layers in a strong uniform perpendicular magnetic field was proposed in Ref.~\cite{ABG}.

\medskip
\par

We present the conditions for  BEC and superfluidity of a two-component weakly interacting Bose gas of dipolar excitons, formed by electron-hole pairs in spatially separated GHAT3 layers. An applied weak magnetic field to this pseudospin-1 monolayer system results in a Zeeman-type splitting of the energy subbands~\cite{BGR}. This dispersion relation consists of three bands: CB, IB and VB. We consider two types of dipolar excitons in double-layer of GHAT3: (a) ``A excitons'', formed as bound states of electrons in CB and holes in IB and (b) ``B excitons'', formed as bound states of electrons in CB and holes in VB. The binding energy of A and B dipolar excitons is calculated. For a two-component weakly interacting Bose gas of dipolar excitons in a GHAT3 double layer, we obtain the energy dispersion of collective excitations, the sound velocity, the superfluid density, and the mean-field critical temperature $T_{c}$ for superfluidity.

\medskip
\par
Our paper is organized in the following way. In Sec. \ref{exciton},  the two-body problem for an electron and a hole, spatially separated in two parallel GHAT3 monolayers, is formulated, and the effective masses and binding energies  are obtained for two types of dipolar excitons. The spectrum of
collective excitations and the sound velocity for the two-component weakly interacting Bose gas of dipolar excitons in the double layer of GHAT3 are derived in Sec.~\ref{collect}. In Sec.~\ref{sup} the superfluidity of the weakly interacting Bose gas of dipolar excitons in the double layer of GHAT3 is  predicted, and the mean-field critical temperature of the phase transition is obtained. The results of our calculations are discussed in Sec.~\ref{disc}.  In Sec.~\ref{conc} our conclusions are reported.

\section{Dipolar excitons in a double layer of $\alpha-T_{3}$ with a mass term}

\label{exciton}

We will consider charge carriers in the conduction band, valence band, and in the intermediate band, which corresponds to the flat band in an $\alpha-T_{3}$ layer without a mass term. In the presence of a weak magnetic field, the low-energy Hamiltonian of the charge carriers in a GHAT3 monolayer at the K and K' points is given by~\cite{BGR}

 \begin{eqnarray}  \label{HamGHAT3}
\hat{\mathcal{H}}_{\lambda} = \left(
\begin{array}{ccc}
\Delta & f (\mathbf{k}) \cos \phi & 0 \\
f^{*} (\mathbf{k}) \cos \phi & 0 & f (\mathbf{k}) \sin \phi  \\
 0 & f^{*} (\mathbf{k}) \sin \phi  & - \Delta %
\end{array}%
\right)  ,
\end{eqnarray}
where the origin in $\mathbf{k}$-space is defined to be around  the K point, $\mathbf{k} = (k_{x}, k_{y})$ and $\tan \theta_{\mathbf{k}} = k_{y}/k_{x}$, $\phi = \tan^{-1} \alpha$, $f (\mathbf{k}) = \hbar v_{F} \left(\lambda k_{x} - i k_{y}\right) = \lambda \hbar v_{F} k e^{-i\lambda \theta_{\mathbf{k}}}$, with $\lambda = \pm 1$ being the valley index at the K and K' points, $2 \Delta$ is the gap  in the energy spectrum of    a GHAT3 layer due to the mass term in the Hamiltonian. In an $\alpha-T_{3}$ layer honeycomb lattice, there is an added fermionic hub atom C at the center of each hexagon. Let the hopping integral be $t_{1}$ between the hub atom and either an A or B atom on the rim and $t_{2}$ between nearest neighbors on the rim  of the hexagon. The ratio of these two nearest neighbor hopping terms is denoted as $t_{2}/t_{1} = \alpha$, where the parameter $\alpha$ satisfies $0 \leq \alpha \leq 1$. The largest value when $\alpha$ is $1$ is for the dice lattice, whereas its value of $0$ corresponds to graphene for decoupled hub from rim atoms~\cite{BGR}.

\medskip
\par

 At small momenta near K and K' points the dispersion for the charge carriers in the conduction band $\epsilon_{CB}(k)$ is given by the relation~\cite{BGR}

\begin{eqnarray}  \label{ecb}
\epsilon_{CB}(k) \approx \Delta + \frac{\hbar^{2} k^{2}}{2 m_{CB}} ,
\end{eqnarray}
where $\mathbf{k} = \mathbf{p}/\hbar$ and $\mathbf{p}$ are the wave vector and momentum of a quasiparticle,  $m_{CB}$ is the effective mass of the charge carriers in the conduction band, given by

\begin{eqnarray}  \label{mCB}
m_{CB} =  \frac{\left(1+ \alpha^{2}\right)\Delta}{2v_{F}^{2}}  ,
\end{eqnarray}
where $v_{F}$ is the Fermi velocity in a GHAT3 layer, and $\varphi = \tan^{-1} \alpha$~\cite{BGR}. At small momenta near K and K' points,  the dispersion for the charge carriers in the valence band $\epsilon_{VB}(k)$ is given by the relation~\cite{BGR}

\begin{eqnarray}  \label{evb}
\epsilon_{VB}(k) \approx - \Delta - \frac{\hbar^{2} k^{2}}{2 m_{VB}} \ ,
\end{eqnarray}
with $m_{VB}$  the effective mass of the charge carriers in the valence band, given by

\begin{eqnarray}  \label{mVB}
m_{VB} =  \frac{\left(1+
\alpha^{2}\right)\Delta}{2v_{F}^{2}\alpha^{2}} .
\end{eqnarray}
At small momenta near K and K' points, the dispersion for the charge carriers   in the intermediate band, corresponding to the flat band in an $\alpha-T_{3}$ layer without a mass term, $\epsilon_{IB}(k)$ is given by the relation~\cite{BGR}

\begin{eqnarray}  \label{eib}
\epsilon_{IB}(k) \approx  - \frac{\hbar^{2} k^{2}}{2 m_{IB}} ,
\end{eqnarray}
where $m_{IB}$ is the effective mass of the charge carriers in the
intermediate band, given by

\begin{eqnarray}  \label{mIB}
m_{IB} =  \frac{\left(1+ \alpha^{2}\right)\Delta}{2v_{F}^{2}\left(1
- \alpha^{2}\right)} .
\end{eqnarray}
It is worthy noting     that there are spin degeneracy and valley degeneracy for the energy of the charge carriers in a GHAT3 layer.

\medskip
\par

\begin{figure} [h]
\includegraphics[width=.60\linewidth]{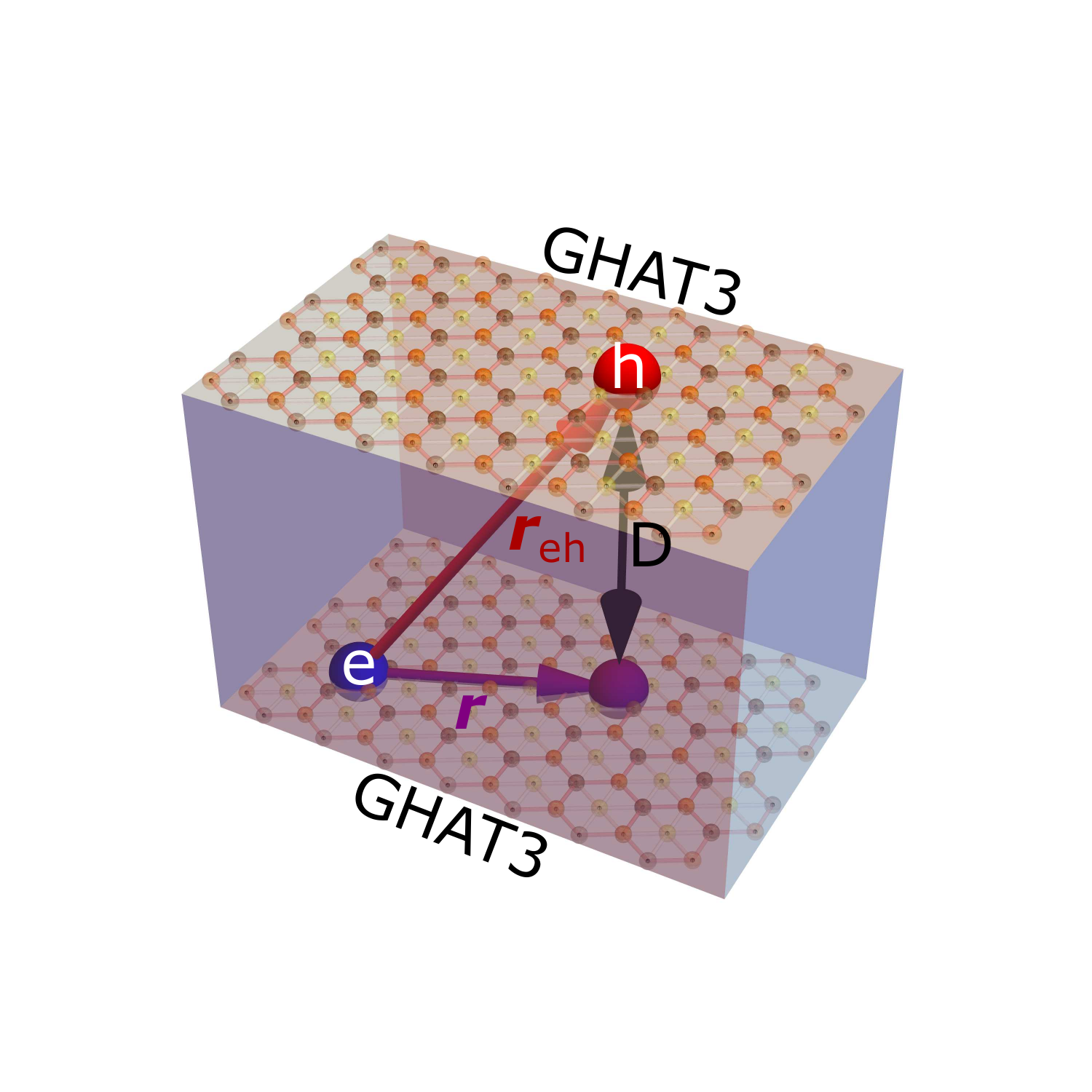}
\caption{(Color online)  Schematic illustration of a dipolar exciton
in a pair of GHAT3 double  layers embedded in an insulating
material.} \label{FIG:1}
\end{figure}

In the system under consideration in this paper, electrons are  confined in a 2D GHAT3 monolayer,
 while an equal number of positive holes are located in a parallel GHAT3 monolayer at a distance $D$ away as demonstrated in Fig.~\ref{FIG:1}.
  This electron-hole system in two parallel GHAT3 layers is treated as a 2D system without interlayer hopping. Due to the absence of  tunneling of electrons and holes between different GHAT3 monolayers,  electron-hole recombination is suppressed by a dielectric barrier with dielectric constant $\epsilon _{d}$ that separates the GHAT3 monolayers. Therefore, the dipolar excitons, formed by electrons and holes, located in two different GHAT3 monolayers, have a longer lifetime than direct excitons. The electron and hole are attracted via electromagnetic interaction
 $ V(r_{eh}),$ where $r_{eh}$ is the distance between the electron and hole, and they could form a bound state, i.e., an exciton, in three-dimensional (3D) space.
 Therefore, in order to determine the binding energy of the exciton  a two body problem in restricted 3D space has to be
 solved.  However, if one projects the electron position vector onto the GHAT3 plane with holes and replaces the relative coordinate vector ${\bf r}_{eh}$ by its
 projection $\mathbf{r}$ on this plane, the potential $V(r_{eh})$ may be expressed as $V(r_{eh})= V(\sqrt{r^{2}+D^{2}}),$ where $r$ is the relative distance between the hole and the projection of the electron position vector onto the GHAT3 plane with holes. A schematic illustration of the dipolar exciton in a GHAT3 double layer is presented in Fig~\ref{FIG:1}. By introducing in-plane coordinates $\mathbf{r}_{1}=(x_{1},y_{1})$ and $\mathbf{r}_{2}=(x_{2},y_{2})$ for the electron and the projection vector of the hole, respectively (where $\mathbf{r}=$ $\mathbf{r}_{1}-\mathbf{r}_{2}$), the dipolar exciton can be described by employing a two-body 2D Schr\"{o}dinger equation with potential $V(\sqrt{r^{2}+D^{2}})$. So that the restricted 3D two-body problem can be reduced to a 2D two-body problem on a GHAT3 layer with the holes.

 The dipolar excitons with spatially separated
electrons and holes in two parallel GHAT3 monolayers can be created
by laser pumping with an applied external voltage. While an electron
in the conduction band and a hole in the valence or intermediate
band are excited due to absorbtion of a photon, voltages are applied
with opposite sign to confine electrons on one layer and holes on
another so  that dipoles point in one direction only.

In our case, ``both'' the energy bands and the exciton modes
referred to the K-point, not one to the $\Gamma$-point and the other
to the K point.  We note that in the dispersion  equations appearing
in Refs.~\cite{BGR,BDGIHR,Malcolm,PhD} the origin of the k-space was
specified to be around the K point, (and not the $\Gamma$ point) as
did several authors investigating $\alpha$-T$_3$. So, our choice of
origin not being the center of the Brillouin zone has precedence.
For graphene,  the plasmon dispersion relation and low-energy bands,
presented by  Ref.~\cite{Wunsch} were both consistently measured
from the K point taken as origin and not the center of the Brillouin
zone.

We consider excitons, formed by the an electron and a hole from the
same valley, because an electron and a hole from different valleys
cannot be excited by absorption of photon due to conservation of
momentum. The reason is that photons carry momenta much smaller than
the difference between K and K' in  reciprocal space.

\medskip
\par

The effective Hamiltonian of an electron and a hole, spatially separated in two parallel GHAT3 monolayers with the interlayer distance $D$ has the following form

\begin{equation}
\hat{H}_{ex}=-\frac{\hbar ^{2}}{2m_{e}}\Delta
_{\mathbf{r}_{1}}-\frac{\hbar ^{2}}{2m_{h}}\Delta _{\mathbf{r}_{2}}
+ V (r)  , \label{rk2}
\end{equation}%
where $\Delta _{\mathbf{r}_{1}}$ and $\Delta _{\mathbf{r}_{2}}$ are the Laplacian operators with respect to the components of the vectors  $\mathbf{r}_{1}$ and  $\mathbf{r}_{2}$, respectively, and $m_{e}$ and $m_{h}$ are the effective masses of the electron and hole, respectively. For CV excitons $m_{e} = m_{CB}$ and $m_{h} = m_{VB}$; and for CI excitons $m_{e} = m_{CB}$ and $m_{h} = m_{IB}$, where $m_{CB}$, $m_{VB}$, and $m_{IB}$ are given by Eqs.~(\ref{mCB}),~(\ref{mVB}), and~(\ref{mIB}), correspondingly. The problem of the in-plane motion of an interacting electron and hole forming the exciton in a GHAT3 double layer can be reduced to that of one particle with the reduced mass $\mu = m_{e}m_{h}/\left(m_{e}+m_{h}\right)$ in a $V(r)$ potential and motion of the center-of-mass of the exciton with the mass $M = m_{e} + m_{h}$. We introduce the coordinates of the center-of-mass $\mathbf{R}$ of an exciton and the coordinate of the relative motion $\mathbf{r}$ of an electron and hole as  $ \mathbf{R} = \left(m_{e}\mathbf{r}_{1} + m_{h}\mathbf{r}_{2}\right)/\left(m_{e}+ m_{h}\right)$ and $\mathbf{r} = \mathbf{r}_{1} - \mathbf{r}_{2}$, correspondingly. The Hamiltonian $\hat{H}_{ex}$ can be represented in the form: $\hat{H}_{ex}= \hat{H}_{R} + \hat{H}_{r}$, where  the Hamiltonians of the motion of the center-of-mass $\hat{H}_{R}$ and relative motion of electron and a hole $\hat{H}_{r}$. The solution of the Schr\"{o}dinger equation for the center-of-mass of an exciton $\hat{H}_{R}\psi (\mathbf{R}) = \mathcal{E}\psi (\mathbf{R})$ is the plane wave $\psi (\mathbf{R}) = e^{i\mathbf{P} \cdot \mathbf{R}/\hbar}$ with the quadratic energy spectrum $\mathcal{E} = P^{2}/(2M)$, where $\mathbf{P}$ is the momentum of the center-of-mass of an exciton.
\medskip
\par

We consider electrons and holes to be located in GHAT3 parallel layers, embedded in a dielectric with the dielectric constant $\epsilon_{d}$.  The potential energy of electron-hole Coulomb attraction is

\begin{eqnarray}\label{el-ho}
V(r) = -\frac{\kappa e^{2}}{\epsilon_{d}\sqrt{r^{2} + D^{2}}}\  ,
\end{eqnarray}
where  $\kappa=9\times 10^{9}\ N\times m^{2}/C^{2}$, $\epsilon_{d}$ is the dielectric constant of the insulator (SiO$_2$ or $h$-BN), surrounding the electron and hole GHAT3 monolayers, forming the double layer. For the $h$-BN barrier we substitute the dielectric constant $\epsilon_{d}= 4.89$, while for the SiO$_2$ barrier we substitute the dielectric constant $\epsilon_{d}=4.5$. For $h$-BN insulating layers, $\epsilon_{d} = 4.89$ is the effective dielectric constant, defined as $\epsilon_{d}  =\sqrt{\varepsilon^{\bot}}\sqrt{\varepsilon^{\parallel}}$~\cite{Fogler}, where $\varepsilon^{\bot}= 6.71$ and $\varepsilon^{\parallel} =3.56$ are the components of the dielectric tensor for $h$-BN~\cite{CaihBN}. Assuming $r\ll D$, we approximate $V(r)$ by the first two terms of the Taylor series and obtain

\begin{eqnarray}  \label{Vap}
V(r)=-V_{0}+\gamma r^{2} , \ \  \text{ where }
V_{0}=\frac{\kappa e^{2}}{\epsilon_{d}D}, \ \ \ \ \  \ \ \ \gamma =\frac{\kappa e^{2}}{%
2\epsilon_{d}D^{3}} .
\end{eqnarray}

\medskip
\par

The solution of the Schr\"{o}dinger equation for the relative motion of an electron and a hole  $\hat{H}_{r}\Psi (\mathbf{r}) = E\Psi (\mathbf{r})$ with  the potential (\ref{Vap}) is reduced to the problem of a 2D harmonic oscillator with the exciton reduced mass $\mu$. Following Refs.~[\onlinecite{Maksym,Iyengar}] one obtains the radial Schr\"{o}dinger equation and the solution for the eigenfunctions for the relative motion of an electron and a hole  in a GHAT3 double layer in terms of associated Laguerre polynomials, which can be written as

\begin{equation}
\Psi_{N L} (\mathbf{r})=\frac{N!}{a^{|L|+1}\sqrt{\widetilde{n}!%
\widetilde{n}^{\prime }!}}2^{-|L|/2}\mathrm{sgn}(L)^{L}r^{|L|}e^{-r^{2}/(4a^{2})}%
\times L_{N}^{|L|}(r^{2}/(2a^{2}))\frac{e^{-iL\varphi }}{(2\pi
)^{1/2}} , \label{rk14}
\end{equation}%
where $N=\mathrm{min}(\widetilde{n},\widetilde{n}^{\prime })$, $L=\widetilde{n}-\widetilde{n}^{\prime}$, $\widetilde{n},$ $\widetilde{n}^{\prime }=0,1,2,3,\ldots $
 are the quantum numbers, $\varphi$ is the polar  angle,
  and $a=\left[ \hbar /\left(2\sqrt{2\mu\gamma }\right)\right]^{1/2}$  is a Bohr radius of a dipolar exciton. The corresponding energy spectrum is given by

\begin{equation}
E_{N L} \equiv E_{e(h)} = - V_{0} + (2N+1+|L|)\hbar \left( \frac{2\gamma}{\mu}%
\right) ^{1/2} .  \label{rk15}
\end{equation}
At the lowest quantum state $N = L = 0$ as it follows from Eq.~(\ref{rk15})  the ground state energy  for the exciton is given by

\begin{equation}
E_{00}=  - V_{0} + \hbar \left( \frac{2\gamma}{\mu}%
\right) ^{1/2} .  \label{rk16}
\end{equation}

\medskip
\par

The important characteristics of the exciton is the square of the in-plane gyration radius $r_{X}^2$. It allows one to estimate the condition when the excitonic gas is dilute enough. One can obtain the square of the in-plane gyration radius $r_{X}$ of a dipolar exciton~\cite{Fogler}, which is the average squared projection of an electron-hole separation onto the plane of a GHAT3 monolayer as

\begin{equation}
r_{X}^{2} \equiv \left\langle r^{2} \right\rangle = \int
\Psi_{00}^{*} (\mathbf{r}) r^{2} \Psi_{00} (\mathbf{r}) d^{2} r =
\frac{2\pi}{2\pi a^{2}}\int_{0}^{+\infty} r^{2} e^{- \frac{r^{2}}{2
a^{2}}} r d r   = 2 a^{2} .
 \label{rx2}
\end{equation}

\medskip
\par

We consider dipolar excitons, formed by an electron in the conduction band and a hole in the valence band (CV excitons) and formed by and electron in the conduction band and a hole in the intermediate valence band (CI excitons). For CV excitons one has

\begin{eqnarray}  \label{mCV}
\mu_{CV} = \frac{m_{CB}m_{VB}}{m_{CB} + m_{VB}} =
\frac{\Delta}{2v_{F}^{2}}; \hspace{2cm} M_{CV} = m_{CB} + m_{VB} =
\frac{\left(1 + \alpha^{2}\right)^{2}\Delta}{2 v_{F}^{2}\alpha^{2}} \  .
\end{eqnarray}
For CI excitons one has

\begin{eqnarray}  \label{mCI}
\mu_{CI} = \frac{m_{CB}m_{IB}}{m_{CB} + m_{IB}} = \frac{\left(1 +
\alpha^{2}\right) \Delta}{2v_{F}^{2}\left(2 - \alpha^{2}\right)};
\hspace{2cm} M_{CI} = m_{CB} + m_{IB} = \frac{\left(1 +
\alpha^{2}\right)\left(2 - \alpha^{2}\right) \Delta}{2
v_{F}^{2}\left(1 - \alpha^{2}\right)}  .
\end{eqnarray}

\section{The collective excitations spectrum and superfluidity for the two-component system of dipolar excitons}
\label{collect}

We consider the dilute limit for dipolar exciton gas in a GHAT3 double layer, when $n_{A}a_{B\ A}^{2}\ll 1$ and $n_{B}a_{B\ B}^{2}\ll 1$, where $n_{A(B)}$ and $a_{B\ A(B)}$ are the concentration and effective exciton Bohr radius for A(B) dipolar excitons, correspondingly.  In the dilute limit, dipolar A and B excitons are formed by electron-hole pairs with the electrons and holes spatially separated in two different GHAT3 layers. We will treat the two-component weakly interacting Bose gas of dipolar excitons in a  GHAT3 double layer by applying the approach analogous to one used  for dipolar excitons in a transition metal dichalcogenide (TMDC) double layer~\cite{BK,BK2}.

\medskip
\par

 Since the dipolar excitons, formed by the charge
carriers in different valleys, are characterized by the same energy,
the exciton states are degenerate with respect to the valley degree
of freedom. Therefore, we consider the Hamiltonian of the weakly
interacting Bose gas of dipolar excitons, formed in a single valley.
We will take into account the degeneracy of the exciton states with
respect to spin and valley degrees of freedom by the introducing the
spin and valley degeneracy factor $s = 16$ below. The Hamiltonian
$\hat{H}$ of the 2D A and B weakly interacting  dipolar excitons can
be written as

\begin{eqnarray}  \label{Ham}
\hat{H}=\hat{H}_{A}+\hat{H}_{B}+\hat{H}_{I}\ ,
\end{eqnarray}
where $\hat{H}_{A(B)}$ are the Hamiltonians of A(B) excitons defined as

\begin{eqnarray}  \label{Ham1}
\hat{H}_{A(B)}=\sum_{\mathbf{k}}E_{A(B)}(k)a_{\mathbf{k}A(B)}^{\dagger }a_{%
\mathbf{k}A(B)}+\frac{g_{AA(BB)}}{2S}\sum_{\mathbf{k}\mathbf{l}\mathbf{m}}a_{%
\mathbf{k}A(B)}^{\dagger }a_{\mathbf{l}A(B)}^{\dagger }a_{A(B)\mathbf{m}%
}a_{A(B)\mathbf{k}+\mathbf{l}-\mathbf{m}}\ ,
\end{eqnarray}
and $\hat{H}_{I}$ is the Hamiltonian of the interaction between A and B excitons presented as

\begin{eqnarray}  \label{Ham2}
\hat{H}_{I}=\frac{g_{AB}}{S}\sum_{\mathbf{k}\mathbf{l}\mathbf{m}}a_{\mathbf{k%
}A}^{\dagger }a_{\mathbf{l}B}^{\dagger }a_{B\mathbf{m}}a_{A\mathbf{k}+%
\mathbf{l}-\mathbf{m}}\ ,
\end{eqnarray}
where $a_{\mathbf{k}A(B)}^{\dagger }$ and $a_{\mathbf{k}A(B)}$ are Bose creation and annihilation operators for A(B) dipolar excitons with the wave vector $\mathbf{k}$, correspondingly, $S$ is the area of the system,  $E_{A(B)}(k)\equiv \epsilon _{A(B)} = \varepsilon _{(0)A(B)}(k)+\mathcal{A}_{A(B)}$ is the energy spectrum of non-interacting A(B) dipolar excitons, respectively,  $ \varepsilon _{(0)A(B)}(k)=\hbar ^{2}k^{2}/(2M_{A(B)})$, $M_{A(B)}$ is an effective mass of non-interacting dipolar excitons, $\mathcal{A}_{A(B)}$ is the constant, which depends  on   A(B) dipolar exciton binding energy and the corresponding gap, $g_{AA(BB)} $ and $g_{AB}$ are the interaction constants for the repulsion between two  A dipolar excitons, two  B dipolar excitons and for the interaction between A and B dipolar excitons, respectively.

\medskip
\par

In dilute system with large interlayer separation $D$, two dipolar excitons, located at distance $R$, repel each other via the dipole-dipole interaction potential $U(R)=\kappa e^{2}D^{2}/(\epsilon _{d}R^{3})$. Following the procedure described in Ref.~\cite{BKKL}, the interaction parameters for the exciton-exciton repulsion in very dilute systems can be obtained implying the exciton-exciton dipole-dipole repulsion exists only at the distances between excitons greater than the distance from the exciton to the classical turning point.
\medskip
\par

The many-particle Hamiltonian for a weakly interacting Bose gas can be diagonalized within the Bogoliubov approximation~\cite{Lifshitz}, replacing the product of four operators in the interaction term by the product of two operators. The Bogoliubov approximation is valid if one assumes  that most of the particles belong to BEC. In this case, in the Hamiltonian one can keep only the terms responsible for the interactions between the condensate and non-condensate particles, while the terms describing the interactions between non-condensate particles, are neglected.
\medskip
\par

Following the procedure, described in Refs.~\cite{BK,BK2}, applying the Bogoliubov approximation~\cite{Lifshitz}, generalized for a two-component weakly interacting Bose gas~\cite{Timmermans,Sun} and introducing the following notation,

\begin{eqnarray}  \label{not}
G_{AA} &=& g_{AA} n_{A} = g n_{A} , \hspace{2cm} G_{BB} = g_{BB}
n_{B} = g n_{B} , \hspace{2cm} G_{AB} = g_{AB}\sqrt{n_{A}n_{B}} = g
\sqrt{n_{A}n_{B}}
\ ,  \nonumber \\
\omega_{A} (k) &=& \sqrt{\varepsilon_{(0)A}^{2}(k) + 2
G_{AA}\varepsilon_{(0)A}(k)} \ ,   \\
\omega_{B}(k) &=& \sqrt{\varepsilon_{(0)B}^{2}(k) + 2
G_{BB}\varepsilon_{(0)B}(k)} \ ,  \nonumber
\end{eqnarray}
one obtains two modes of the spectrum of Bose collective excitations $\varepsilon_{j}(k)$

\begin{eqnarray}  \label{col}
\varepsilon_{j}(k) = \sqrt{\frac{\omega_{A}^{2}(k) +
\omega_{B}^{2}(k) + (-1)^{j-1}\sqrt{\left(\omega_{A}^{2}(k) -
\omega_{B}^{2}(k)\right)^{2} +
\left(4G_{AB}\right)^{2}\varepsilon_{(0)A}(k)\varepsilon_{(0)B}(k)}
}{2}} \ ,
\end{eqnarray}
where $j=1$, $2$. In our approach, the condition $G_{AB}^{2} =
G_{AA}G_{BB}$ holds.
\medskip
\par

At small momenta $p = \hbar k$, when $\varepsilon_{(0)A}(k) \ll G_{AA}$ and $\varepsilon_{(0)B}(k) \ll G_{BB}$,  expanding the spectrum of collective excitations $ \varepsilon_{j}(k)$  up to the first order with respect to the momentum $p$, one obtains two sound modes in the spectrum of the collective excitations $\varepsilon_{j}(p) = c_{j}p$, where $c_{j}$ is the sound velocity written as

\begin{eqnarray}  \label{c}
c_{j} = \sqrt{\frac{G_{AA}}{2M_{A}} + \frac{G_{BB}}{2M_{B}} + (-1)^{j-1}%
\sqrt{\left(\frac{G_{AA}}{2M_{A}} - \frac{G_{BB}}{2M_{B}}\right)^{2} + \frac{%
G_{AB}^{2}}{M_{A}M_{B}}}} \ ,
\end{eqnarray}
 At $j = 1$, the  spectrum of collective excitations is determined by the non-zero sound velocity $c_{1}$, while at $j = 2$ the sound velocity vanishes with $c_{2} = 0$. At large momenta, for the conditions when $\varepsilon _{(0)A}(k)\gg G_{AA}$ and $ \varepsilon _{(0)B}(k)\gg G_{BB}$, one obtains two parabolic modes of collective excitations with the spectra $\varepsilon _{1}(k)=\varepsilon _{(0)A}(k)$ and $\varepsilon _{2}(k)=\varepsilon _{(0)B}(k)$, if $M_{A}<M_{B}$ and if $ M_{A}>M_{B}$ with the spectra $\varepsilon _{1}(k)=\varepsilon _{(0)B}(k)$ and $\varepsilon _{2}(k)=\varepsilon _{(0)A}(k)$.

\section{Superfluidity of the weakly-interacting Bose gas of dipolar excitons}
\label{sup}

Since when $j = 2$ the sound velocity vanishes, below we take into
account only the branch of the spectrum of collective excitations at
$j = 1$, neglecting the branch at $j = 2$.  According to
Refs.~\cite{Lifshitz,Abrikosov}, it is clear that we need a finite
sound velocity for superfluidity. Since the branch of the collective
excitations at zero sound velocity for the collective excitations
corresponds to the zero energy of the quasiparticles (which means
that  no quasiparticles are created with zero sound velocity), this
branch does not lead to the dissipation of energy resulting in
finite viscosity and, therefore, does not influence the Landau
critical velocity. This is the reason for eliminating the zero sound
velocity case in our considerations here. The weakly-interacting gas
of dipolar excitons in the double layer of GHAT3 satisfies the
Landau criterion for superfluidity~\cite{Lifshitz,Abrikosov},
because at small momenta the energy spectrum of the quasiparticles
in  the weakly-interacting gas of dipolar excitons  at $j = 1$ is
sound-like with the finite sound velocity $c_{1}$. In the moving
weakly-interacting gas of dipolar excitons the quasiparticles are
created at velocities above the velocity of sound, and the critical
velocity for  superfluidity reads as $v_{c} = c_{1}$. The difference
between the ideal Bose gas and two-component  weakly interacting
Bose gas of dipolar excitons is that while the spectrum of ideal
Bose gas has no branch with finite sound velocity, the dipolar
exciton system under consideration has one branch in the spectrum of
collective excitations with  finite sound velocity at $j = 1$ due to
exciton-exciton interaction. Therefore, at low temperatures, the
two-component system of dipolar excitons exhibits superfluidity due
to exciton-exciton interactions, while the ideal Bose gas does not
demonstrate superfluidity.

\medskip
\par

We defined the density of the superfluid component $\rho _{s}(T)$ as $\rho _{s}(T)=\rho -\rho _{n}(T)$, where $\rho =M_{A}n_{A}+M_{B}n_{B}$ is the total 2D density of the dipolar excitons and $\rho _{n}(T)$ denotes the density of the normal component. The density $\rho _{n}(T)$ of the normal component can be defined using standard procedure \cite{Pitaevskii}. The assumption that the dipolar exciton system moves with a velocity $\mathbf{u}$ implies that the superfluid component moves with the velocity $\mathbf{u}$. The energy dissipation at nonzero temperatures $T$ is characterized by the occupancy of quasiparticles  in this system. Since the density of quasiparticles is small at low temperatures, the gas of quasiparticles can be treated as an ideal Bose gas. In order to obtain the density of the  superfluid component, one can define the total mass flow for a Bose gas of quasiparticles in the frame, in which the superfluid component is assumed to be at rest, as

\begin{eqnarray}  \label{nnor}
\mathbf{J}= s \int \frac{d^{2}p}{(2\pi \hbar )^{2}}\mathbf{p}
f\left[ \varepsilon _{1}(p)-\mathbf{p}\cdot\mathbf{u}\right]  \ ,
\end{eqnarray}
where $s = 16$ is the spin and valley degeneracy factor,  $f\left[
\varepsilon _{1}(p))\right] =\left( \exp \left[ \varepsilon
_{1}(p)/(k_{B}T)\right] -1\right) ^{-1}$ is the Bose-Einstein
distribution function for the quasiparticles with the dispersion
$\varepsilon _{1}(p)$, and $k_{B}$ is the Boltzmann constant.
Expanding the expression under the integral in Eq.~(\ref{nnor}) up
to the first order  with respect to
$\mathbf{p}\cdot\mathbf{u}/(k_{B}T)$, one has:

\begin{eqnarray}  \label{J_Tot}
\mathbf{J}=- s \frac{\mathbf{u}}{2}\int \frac{d^{2}p}{(2\pi \hbar )^{2}}%
p^{2} \frac{\partial f\left[ \varepsilon _{1}(p)\right] }{\partial
\varepsilon _{1}(p)} \ .
\end{eqnarray}
The density $\rho _{n}$ of the normal component in the moving
weakly-interacting Bose gas of dipolar excitons is defined
as~\cite{Pitaevskii}

\begin{eqnarray}  \label{J_M}
\mathbf{J}=\rho _{n}\mathbf{u}\ .
\end{eqnarray}
Employing Eqs.~(\ref{J_Tot}) and~(\ref{J_M}), one derives  the
normal component density as

\begin{eqnarray}  \label{rhon}
\rho _{n}(T)=-\frac{s}{2}\int \frac{d^{2}p}{(2\pi \hbar
)^{2}}p^{2}\frac{\partial f\left[ \varepsilon _{1}(p)\right]
}{\partial \varepsilon _{1}(p)} \ .
\end{eqnarray}
\medskip
\par

At low temperatures $k_{B}T\ll M_{A(B)}c_{j}^{2}$, the small momenta ($\varepsilon _{(0)A}(k)\ll G_{AA}$ and $ \varepsilon _{(0)B}(k)\ll G_{BB}$) make the dominant contribution to the integral on the right-hand side of Eq.~(\ref{rhon}). The quasiparticles with such small momenta are characterized by the sound spectrum $\varepsilon _{1}(k)=c_{1}k$ with the sound velocity defined  by Eq.~(\ref{c}). By substituting $\varepsilon _{1}(k)=c_{1}k$ into Eq.~(\ref{rhon}),   we obtain

\begin{eqnarray}  \label{rhon1}
\rho _{n}(T)=\frac{3s \zeta (3)}{2\pi \hbar
^{2}c_{1}^{4}}k_{B}^{3}T^{3} \ ,
\end{eqnarray}
where $\zeta (z)$ is the Riemann zeta function ($\zeta (3)\simeq
1.202$).

\medskip
\par

The mean field critical temperature $T_{c}$ of the phase transition at which the superfluidity occurs, implying neglecting  the interaction between the quasiparticles, is obtained from the condition $%
\rho_{s}(T_{c}) = 0$~\cite{Pitaevskii}:

\begin{eqnarray}  \label{Tc1}
\rho_{n} (T_{c}) = \rho = M_{A} n_{A} + M_{B}n_{B} \ .
\end{eqnarray}
At low  temperatures $k_{B}T\ll M_{A(B)}c_{1}^{2}$ by substituting Eq.~(\ref%
{rhon1}) into Eq.~(\ref{Tc1}), one derives

\begin{eqnarray}  \label{Tc2}
T_{c}=\left[ \frac{2\pi \hbar ^{2}\rho c_{1}^{4}}{3\zeta (3)s
k_{B}^{3}}\right] ^{1/3}\ .
\end{eqnarray}

 While Bose-Einstein condensation occurs at absolute
zero even in a two-dimensional (2D) system, it is well known that in
a 2D bosonic system, Bose-Einstein condensation does not occur at
finite temperature, and only the quasi-long-range order appears. In
this paper, we have obtained the mean field critical temperature
$T_{c}$ of the phase transition at which  superfluidity appears
without claiming BEC in a 2D system at finite temperature. In this
work, we have considered BEC only at absolute zero temperature.

\section{Discussion}
\label{disc}

\begin{figure}[h!]
\centering
\includegraphics[width=0.6\textwidth]{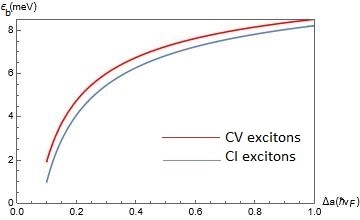}
\caption{(Color online) \  The exciton binding energy
$\mathcal{E}_{b}(\alpha,\Delta,D)$ for CV and CI excitons as
functions of the gap $\Delta$ for chosen parameter $\alpha = 0.6$ and
interlayer separations $D = 25 \ \mathrm{nm}$. The lattice constant
of $\alpha-T_{3}$ is $a = 2.46 \ {\AA}$. } \label{Fig:2}
\end{figure}

In this section we now  discuss the results of our calculations. In Fig.~\ref{Fig:2}, we present the results for the exciton binding energy $\mathcal{E}_{b}(\alpha,\Delta,D)$ for CV and CI excitons as functions of the gap $\Delta$ for chosen parameter $\alpha = 0.6$ and interlayer separations $D = 25 \ \mathrm{nm}$. According to Fig.~\ref{Fig:2}, $\mathcal{E}_{b}(\alpha,\Delta,D)$ is an increasing function  of $\Delta$, whereas for CV excitons the exciton binding energy is slightly larger than that for CI excitons.

\medskip
\par

\begin{figure}[h!]
\centering
\includegraphics[width=0.6\textwidth]{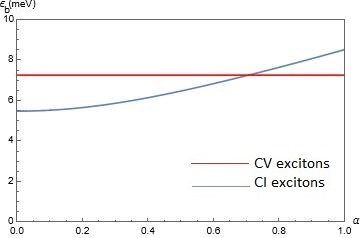}
\caption{(Color online) \  The exciton binding energy
$\mathcal{E}_{b}(\alpha,\Delta,D)$ for CV and CI excitons as
functions of the parameter $\alpha$ for chosen gap $\Delta = 0.5
\hbar v_F/a $ and interlayer separations $D = 25 \ \mathrm{nm}$.}
\label{Fig:3}
\end{figure}

In Fig.~\ref{Fig:3}, we present our results  for the exciton binding energy $\mathcal{E}_{b}(\alpha,\Delta,D)$ for CV and CI excitons as functions of the parameter $\alpha$ for chosen gap $\Delta = 0.5 \hbar v_F/a $ and interlayer separations $D = 25 \ \mathrm{nm}$.  According to Fig.~\ref{Fig:3}, $\mathcal{E}_{b}(\alpha,\Delta,D)$ does not depend on $\alpha$ for CV excitons, whereas it is an increasing function of $\alpha$ for CI excitons.  At $\alpha \lesssim 0.7$ $\mathcal{E}_{b}(\alpha,\Delta,D)$ for CV excitons is larger than for CI excitons, while at $\alpha \gtrsim 0.7$ $\mathcal{E}_{b}(\alpha,\Delta,D)$ for CI excitons is larger than for CV excitons.

\medskip
\par

\begin{figure}[h!]
\centering
\includegraphics[width=0.6\textwidth]{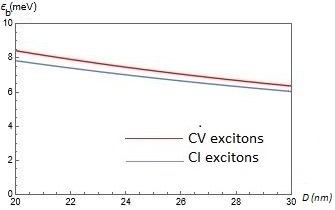}
\caption{(Color online) \  The exciton binding energy
$\mathcal{E}_{b}(\alpha,\Delta,D)$ for CV and CI excitons as
functions of the interlayer separation $D$ for chosen parameter
$\alpha = 0.6$ and gap $\Delta = 0.5 \hbar v_F/a$.} \label{Fig:4}
\end{figure}

 In Fig.~\ref{Fig:4}, we present the results of our calculations for the exciton binding energy $\mathcal{E}_{b}(\alpha,\Delta,D)$ for CV and CI excitons  as   functions of the interlayer separation $D$ for chosen parameter $\alpha = 0.6$ and gap $\Delta = 0.5 \hbar v_F/a$.  According to Fig.~\ref{Fig:4}, $\mathcal{E}_{b}(\alpha,\Delta,D)$ is a decreasing function of $D$, whereas for CV excitons the exciton binding energy is slightly larger than for CI excitons.

\medskip
\par

\begin{figure}[h!]
\centering
\includegraphics[width=0.4\textwidth]{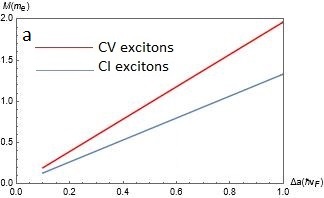}
\includegraphics[width=0.4\textwidth]{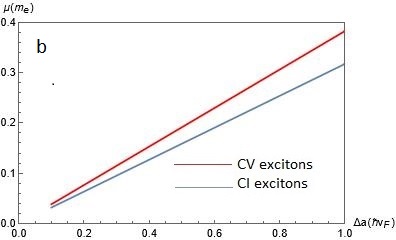}
\caption{(Color online)\  The effective masses of a dipolar exciton
for CV and CI excitons as functions of the gap $\Delta$ for chosen
$\alpha = 0.6$ for (a) center-of-mass exciton mass $M$ on
the left panel and (b) reduced exciton mass $\mu$, on the
right.} \label{Fig:5}
\end{figure}

In Fig.~\ref{Fig:5}, we present plots of the effective masses for CV and CI dipolar excitons as functions of the gap $\Delta$ for chosen $\alpha = 0.6$ for (a) center-of-mass exciton mass $M$ on the left-hand side and (b) reduced exciton mass $\mu$, on the right. According to Fig.~\ref{Fig:5}, both $M$ and $\mu$ for the CV and CI excitons are increasing functions of $\Delta$, while for CV excitons both $M$ and $\mu$  are slightly larger than for CI excitons.

\medskip
\par

\begin{figure}[h!]
\centering
\includegraphics[width=0.4\textwidth]{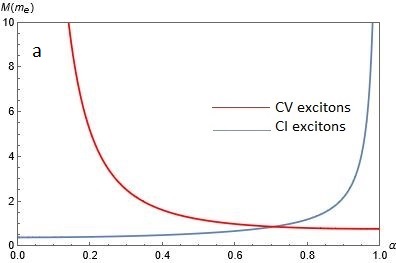}
\includegraphics[width=0.4\textwidth]{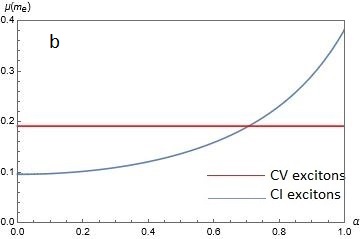}
\caption{(Color online)\  The effective masses of dipolar excitons
for CV and CI excitons as functions of  the hopping parameter $\alpha$  for
chosen gap $\Delta  =0.5 \hbar v_F/a$ for (a) center-of-mass exciton
mass $M$ in the left panel and (b) reduced exciton mass $\mu$,
on the right.}
\label{Fig:6}
\end{figure}

Figure~\ref{Fig:6} shows the effective masses of a dipolar exciton for CV and CI excitons as functions of $\alpha$  for chosen $\Delta  =0.5 \hbar v_F/a$ for (a) center-of-mass exciton mass $M$ in the left panel and (b) reduced exciton mass $\mu$, on the right. According to Fig.~\ref{Fig:6}, for CV excitons $M$ is a decreasing function of $\alpha$, whereas $\mu$ does not depend on $\alpha$. For CI excitons,  both $M$ and $\mu$ increase  as $\alpha$ is  increased.   For $\alpha \lesssim 0.7$, both $M$ and $\mu$ for CV excitons are larger than for CI excitons, but when $\alpha \gtrsim 0.7$ $\mathcal{E}_{b}(\alpha,\Delta,D)$  both $M$ and $\mu$ for CV excitons are smaller than for CI excitons.

\medskip
\par

\begin{figure}[h!]
\centering
\includegraphics[width=0.6\textwidth]{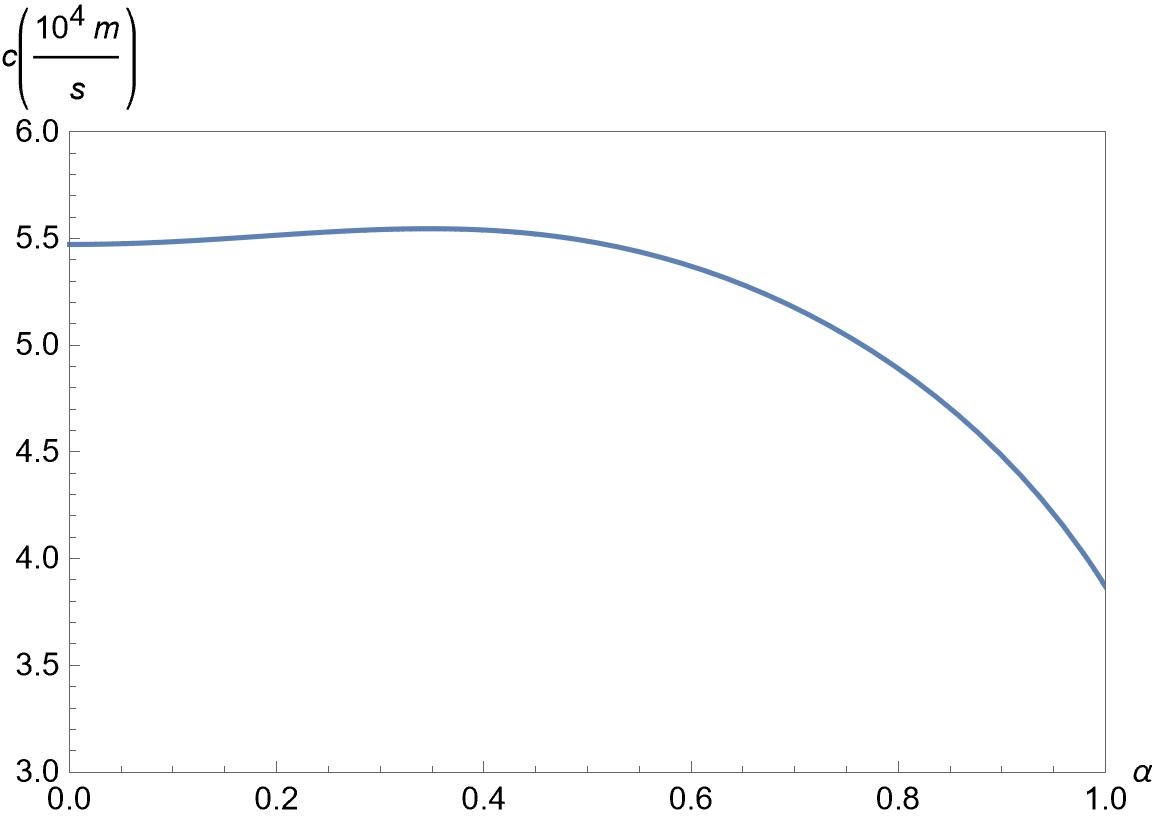}
\caption{(Color online) Plot of the sound velocity $c \equiv c_{1}$
 versus  $\alpha$  for chosen gap $\Delta =
\hbar v_F /a$, interlayer separations $D = 25 \  \mathrm{nm}$ at
fixed concentrations $n_{A}=50\times 10^{11} \ \mathrm{cm^{-2}}$ and
$n_{B}=50\times 10^{11} \ \mathrm{cm^{-2}}$ of A and B excitons,
respectively.}
\label{Fig:7}
\end{figure}

Figure~\ref{Fig:7} demonstrates  the  dependence of the sound velocity $c \equiv c_{1}$  on the hopping parameter $\alpha$  for chosen $\Delta = \hbar v_F /a$, interlayer separations $D = 25 \ \mathrm{nm}$ at fixed concentrations $n_{A}=50\times 10^{11} \ \mathrm{cm^{-2}}$ and $n_{B}=50\times 10^{11} \ \mathrm{cm^{-2}}$ of A and B excitons, respectively. According to Fig.~\ref{Fig:7}, $c$ does not depend much on $\alpha$ when $\alpha \lesssim 0.5$, while for $\alpha \gtrsim 0.5$, the sound velocity $c$ is a decreasing function of $\alpha$.

\medskip
\par

\begin{figure}[h!]
\centering
\includegraphics[width=0.6\textwidth]{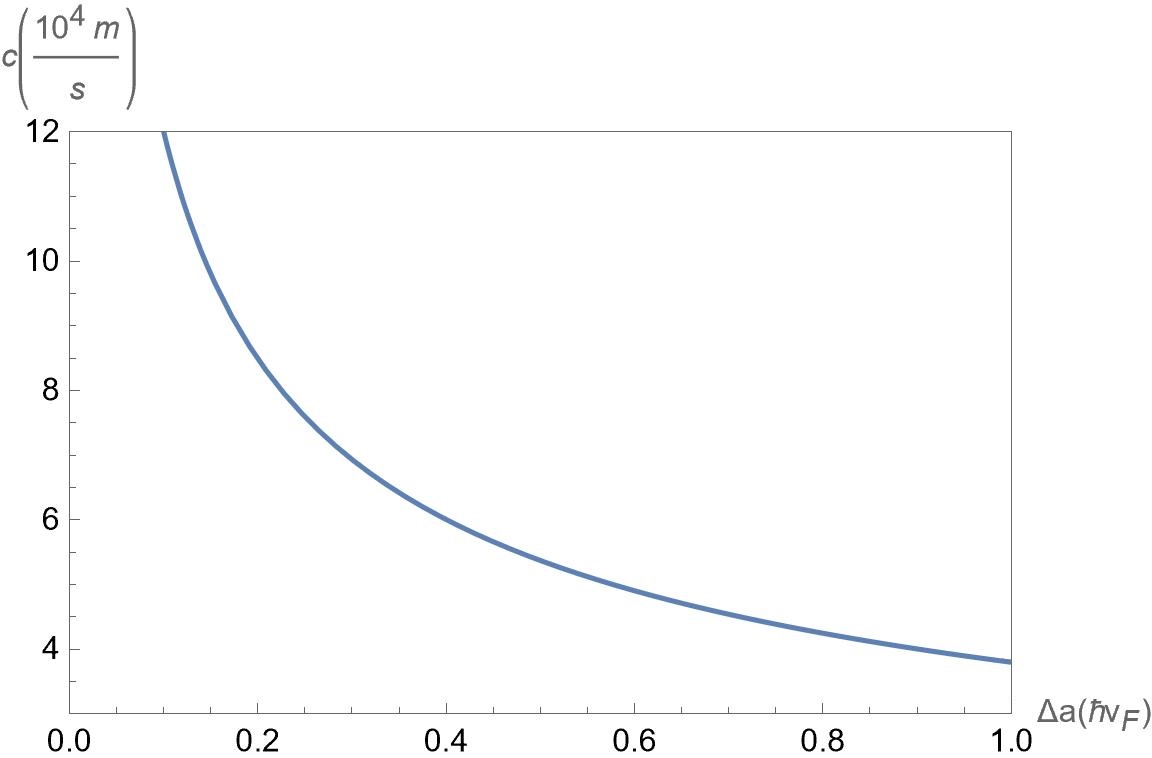}
\caption{(Color online) The sound velocity $c \equiv c_{1}$
 versus the gap $\Delta$    for chosen parameter
$\alpha$=0.6, interlayer separations $D=25 \ \mathrm{nm}$ at the
fixed concentrations $n_{A}=50\times 10^{11} \ \mathrm{cm^{-2}}$ and
$n_{B}=50\times 10^{11} \ \mathrm{cm^{-2}}$ of A and B excitons,
respectively.}
\label{Fig:8}
\end{figure}

In Fig.~\ref{Fig:8}, we plot the  sound velocity $c \equiv c_{1}$ versus the gap $\Delta$    for chosen parameter $\alpha = 0.6$, interlayer separations $D=25 \ \mathrm{nm}$ for  chosen concentrations $n_{A}=50\times 10^{11} \ \mathrm{cm^{-2}}$ and $n_{B}=50\times 10^{11} \ \mathrm{cm^{-2}}$ of A and B excitons, respectively.  According to Fig.~\ref{Fig:8}, the sound velocity $c$ is a decreasing function of  $\Delta$.

\medskip
\par

\begin{figure}[h!]
\centering
\includegraphics[width=0.6\textwidth]{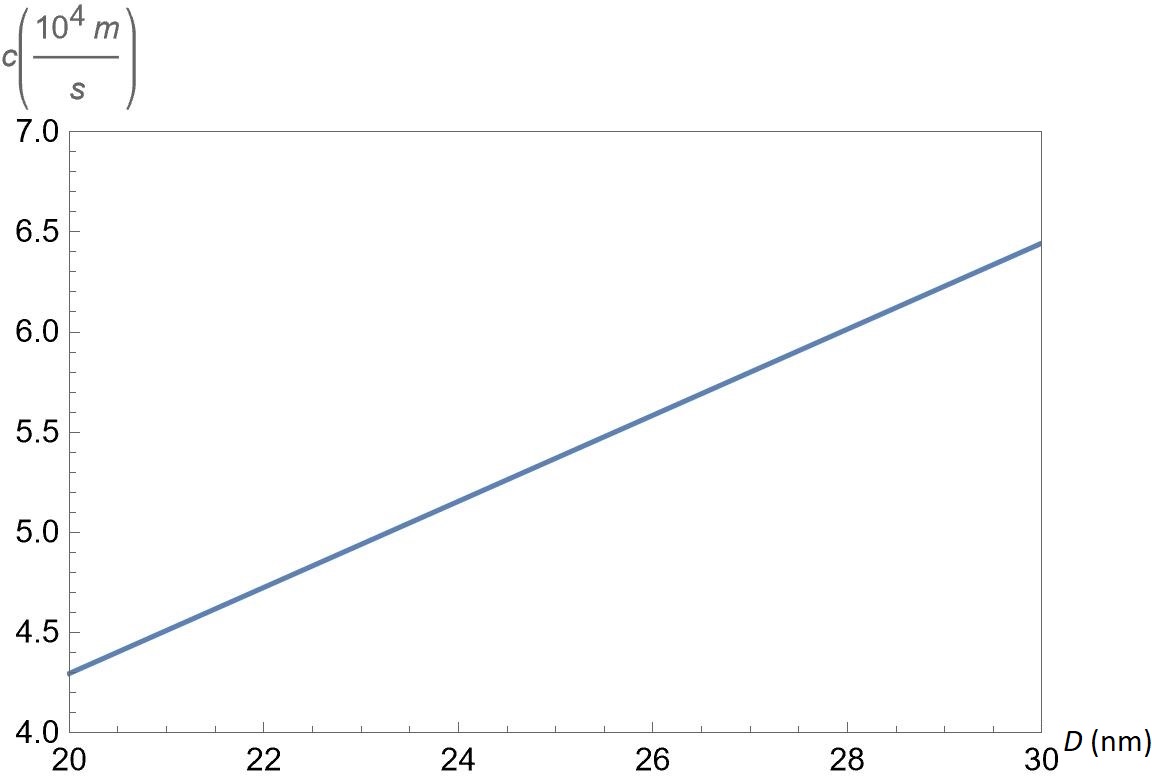}
\caption{(Color online) The sound velocity $c \equiv c_{1}$ versus
the interlayer separation $D$   for chosen parameter $\alpha=0.6$
and gap $\Delta=0.5 \hbar v_F/a$, at fixed concentrations
$n_{A}=50\times 10^{11}\ \mathrm{cm^{-2}}$ and $n_{B}=50\times 10^{11} \ \mathrm{cm^{-2}}$ of A and B excitons, respectively.}
\label{Fig:9}
\end{figure}

In Fig.~\ref{Fig:9}, we show  the  sound velocity $c \equiv c_{1}$ as a function of the interlayer separation $D$   for hopping parameter $\alpha=0.6$ and gap $\Delta=0.5 \hbar v_F/a$, for fixed concentrations $n_{A}=50\times 10^{11}\ \mathrm{cm^{-2}}$ and $n_{B}=50\times 10^{11} \ \mathrm{cm^{-2}}$ of A and B excitons, respectively. According to Fig.~\ref{Fig:9}, the sound velocity $c$ is an increasing function of  $D$.

\medskip
\par

\begin{figure}[h!]
\centering
\includegraphics[width=0.6\textwidth]{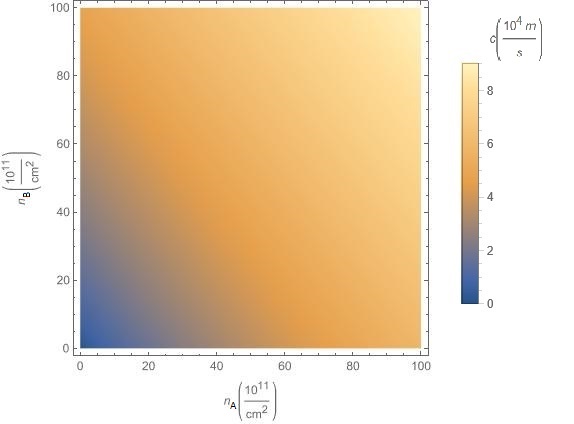}
\caption{(Color online) The sound velocity $c \equiv c_{1}$ versus
the concentrations $n_{A}$ and $n_{B}$ of A and B excitons,
respectively, for chosen parameter $\alpha=0.6$ and gap $\Delta=0.5 \hbar v_F/a$, at the fixed interlayer separation $D= 25 \ \mathrm{nm}$.}
\label{Fig:10}
\end{figure}

In Fig.~\ref{Fig:10}, we illustrate the dependence of the sound velocity $c \equiv c_{1}$  on the concentrations $n_{A}$ and $n_{B}$ of A and B excitons, respectively for chosen hopping parameter $\alpha=0.6$ and gap $\Delta=0.5 \hbar v_F/a$, at  fixed interlayer separation $D= 25 \ \mathrm{nm}$. According to Fig.~\ref{Fig:10}, the sound velocity $c$ is an increasing function of  both concentrations $n_{A}$ and $n_{B}$.

\medskip
\par

\begin{figure}[h!]
\centering
\includegraphics[width=0.6\textwidth]{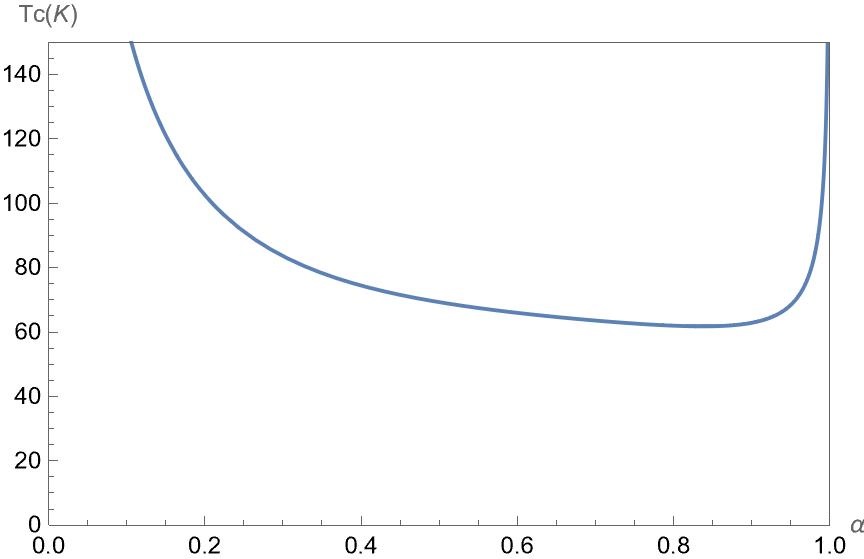}
\caption{(Color online) The mean-field  phase transition critical
temperature $T_{c}(n_{A}, n_{B},\alpha, \Delta, D)$  versus   the
parameter $\alpha$  for chosen gap $\Delta=0.5 \hbar v_F/a$,
interlayer separations $D=25 \ \mathrm{nm}$ at the fixed
concentrations $n_{A}=50\times 10^{11} \ \mathrm{cm^{-2}}$ and
$n_{B}=50\times 10^{11} \ \mathrm{cm^{-2}}$ of A and B excitons,
respectively.}
\label{Fig:11}
\end{figure}

In Fig.~\ref{Fig:11}, we present  the   mean-field  phase transition critical temperature $T_{c}(n_{A}, n_{B},\alpha, \Delta, D)$  as a function of the parameter $\alpha$  for chosen gap $\Delta=0.5 \hbar v_F/a$, interlayer separations $D=25 \ \mathrm{nm}$ at the fixed concentrations $n_{A}=50\times 10^{11} \ \mathrm{cm^{-2}}$ and $n_{B}=50\times 10^{11} \ \mathrm{cm^{-2}}$ of A and B excitons, respectively.   According to Fig.~\ref{Fig:11}, $T_{c}$ is a decreasing function of $\alpha$ at $\alpha \lesssim 0.9$, while at $\alpha \gtrsim 0.9$ the critical temperature $T_{c}$ is an increasing function of $\alpha$.

\medskip
\par

\begin{figure}[h!]
\centering
\includegraphics[width=0.6\textwidth]{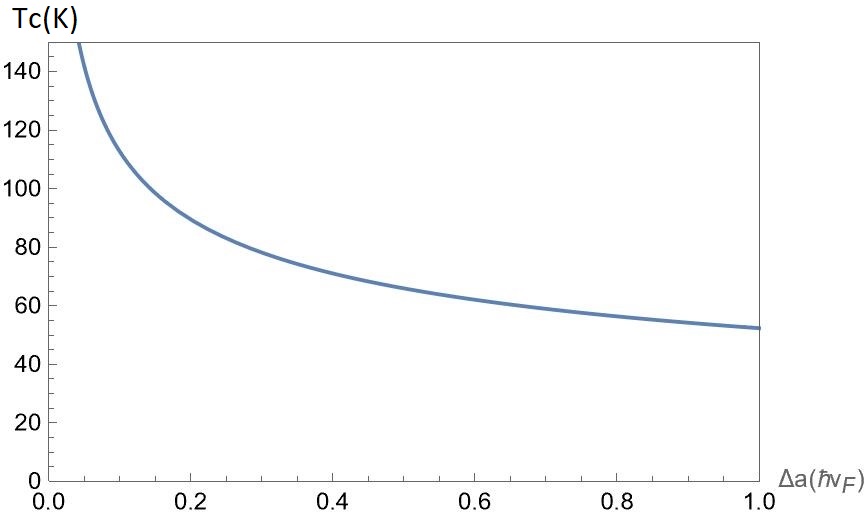}
\caption{(Color online) The mean-field  phase transition critical
temperature $T_{c}(n_{A}, n_{B},\alpha, \Delta, D)$  versus the gap
$\Delta$    for chosen parameter $\alpha=0.6$, interlayer
separations $D=25 \ \mathrm{nm}$ at the fixed concentrations
$n_{A}=50\times 10^{11} \ \mathrm{cm^{-2}}$ and $n_{B}=50\times 10^{11} \ \mathrm{cm^{-2}}$ of A and B excitons, respectively.}
\label{Fig:12}
\end{figure}

In Fig.~\ref{Fig:12}, we present the   mean-field  phase transition critical temperature $T_{c}(n_{A}, n_{B},\alpha, \Delta, D)$  as a function of the gap $\Delta$    for chosen parameter $\alpha=0.6$, interlayer separations $D=25 \ \mathrm{nm}$ at the fixed concentrations $n_{A}=50\times 10^{11} \ \mathrm{cm^{-2}}$ and $n_{B}=50\times 10^{11} \ \mathrm{cm^{-2}}$ of A and B excitons, respectively.   According to Fig.~\ref{Fig:12}, the criticaltemperature $T_{c}$ is a decreasing function of $\Delta$.

\medskip
\par

\begin{figure}[h!]
\centering
\includegraphics[width=0.6\textwidth]{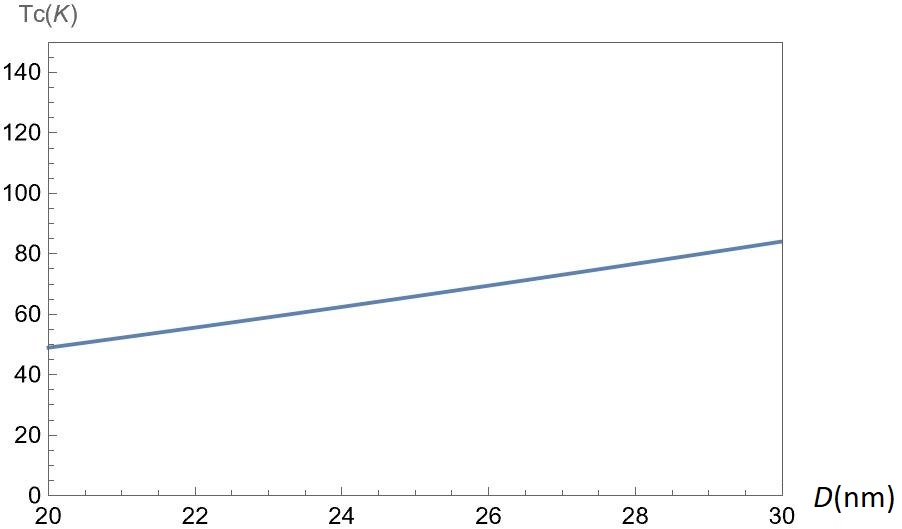}
\caption{(Color online) The mean-field  phase transition critical
temperature $T_{c}(n_{A}, n_{B},\alpha, \Delta, D)$  versus the
interlayer separation $D$     for chosen parameter $\alpha=0.6$ and gap $\Delta=0.5 \hbar v_F/a$, at fixed concentrations
$n_{A}=50\times 10^{11} \ \mathrm{cm^{-2}}$ and $n_{B}=50\times 10^{11} \ \mathrm{cm^{-2}}$ of A and B excitons, respectively.}
\label{Fig:13}
\end{figure}

In Fig.~\ref{Fig:13}, we demonstrate  the   mean-field  phase transition critical temperature $T_{c}(n_{A}, n_{B},\alpha, \Delta, D)$  as a function of the interlayer separation $D$     for chosen parameter $\alpha=0.6$ and gap $\Delta=0.5 \hbar v_F/a$, at the fixed concentrations $n_{A}=50\times 10^{11} \ \mathrm{cm^{-2}}$ and $n_{B}=50\times 10^{11} \ \mathrm{cm^{-2}}$ of A and B excitons, respectively. According to Fig.~\ref{Fig:13},  the critical temperature $T_{c}$ is an increasing function of $D$.

\medskip
\par

\begin{figure}[h!]
\centering
\includegraphics[width=0.6\textwidth]{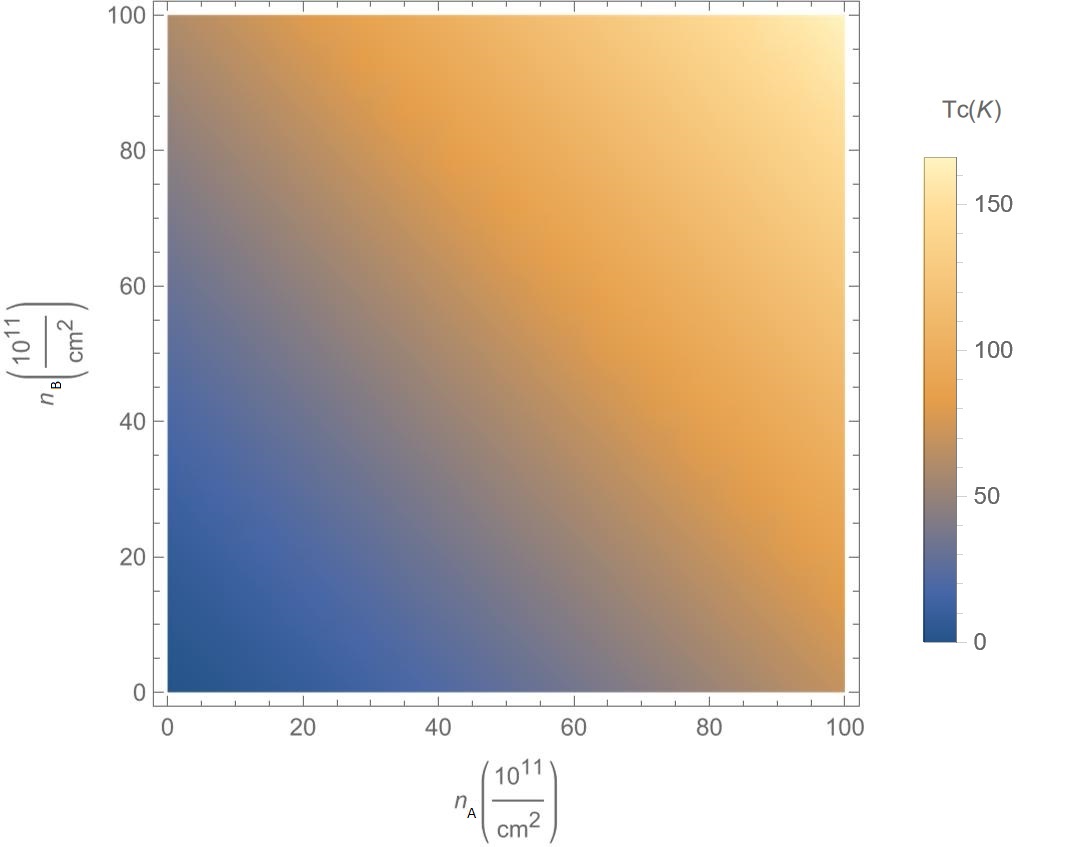}
\caption{(Color online) Density plot for the mean-field  phase transition critical temperature $T_{c}(n_{A}, n_{B},\alpha, \Delta, D)$  versus the concentrations $n_{A}$ and $n_{B}$ of A and B excitons, respectively, for chosen parameter $\alpha=0.6$ and gap $\Delta=0.5 \hbar v_F/a$, at the fixed interlayer separation $D=25 \ \mathrm{nm}$.}
 \label{Fig:14}
\end{figure}

In Fig.~\ref{Fig:14}, we present  density plots for the mean-field  phase transition critical temperature $T_{c}(n_{A}, n_{B},\alpha, \Delta, D)$  as a function of the concentrations $n_{A}$ and $n_{B}$ of A and B excitons, respectively for chosen parameter $\alpha=0.6$ and gap $\Delta=0.5 \hbar v_F/a$, at the fixed interlayer separation $D=25 \ \mathrm{nm}$. According to Fig.~\ref{Fig:14}, the  the critical temperature $T_{c}$ is an increasing function of both the concentrations $n_{A}$ and $n_{B}$.

 At a formal level, the weakly interacting Bose gas of A
and B dipolar excitons in a GHAT3 double layer is similar to the
two-component weakly interacting Bose gas of trapped cold atoms in a
planar harmonic trap. The spectrum of collective excitations in the
Bogoliubov approximation for dipolar excitons in a GHAT3 double
layer is similar to one for a two-component BEC of trapped cold
atoms, studied in Refs~\cite{Timmermans,Sun}.

The gap parameter $\Delta$, has a dual role, since it appears as
chemical potential in the Hamltonian, as also in the mass of the
excitons through the band curvature. According to Figs.~\ref{Fig:2}
and~\ref{Fig:12}, the dipolar exciton binding energy is an
increasing function of the gap $\Delta$, while is the mean field
phase transition temperature $T_{c}$ is a decreasing function of the
gap $\Delta$. Therefore, there should be an optimal value for
$\Delta$, which would correspond to relatively high $T_{c}$ at the
relatively high dipolar exciton binding energy. The latter condition
provides formation of the superfluid phase by the relatively stable
dipolar excitons.

Note that electron-hole superfluids can be formed not only in the
BEC regime but also in the BCS-BEC crossover regime~\cite{Hamilton}.
Quantum Monte Carlo simulations analyzing the BCS-BEC crossover
regime for electron-hole systems have been performed~\cite{Lopez}.
In this Paper we concentrate on dilute electron-hole system, which
corresponds to the BEC, which matches experimentally achievable
densities in the electron-hole systems in 2D materials. BCS regime
requires higher concentrations beyond the model of weakly
interacting Bose gas. The studies of BCS regime and BEC-BCS
crossover for an electron-hole superfluid in a GHAT3 double layer
seem to be promising direction for future studies.

The considered system of dipolar excitons in a GHAT3 double layer
has also a strong similarity, with photon condensation in a cavity,
The collective modes and possibility of Kosterlitz-Thouless phase
transition to the superfluid phase~\cite{Kosterlitz} has been
studied for a photon condensation in a cavity in Ref.~\cite{Vyas}.
If we consider only one type of excitons in a GHAT3 double layer,
assuming the concentration of the excitons of another type to be
zero, the expressions for the spectrum of collective excitations
reported in this Paper can be reduced to the expressions similar to
Ref.~\cite{Vyas}.

The Kosterlitz-Thouless phase transition to the superfluid
phase~\cite{Kosterlitz} can be inferred from the variation of the
superfluid density, which has been computed in this Paper.

Note that in this Paper we did not consider vortices, as within the
mean field approximation it was assumed that the number of
quasiparticles is relatively. However, beyond the mean field
approximation it is possible to consider the properties of vortices
in the system of dipolar excitons. Thus, the dynamical creation of
fractionalized vortices and vortex lattices can be considered by
applying the approach, developed for the BEC of cold atoms in
Ref.~\cite{Ji}.

The Josephson phenomena for two trapped condensates of dipolar
excitons can be studied by applying the approach similar to the one,
developed for non-Abelian Josephson effect between two $F=2$ spinor
Bose-Einstein condensates of cold atoms in double optical
traps~\cite{Qi}.

\medskip
\par

\section{Conclusions}
\label{conc}

This paper is devoted to an investigation of  the existence of BEC and superfluidity of dipolar excitons in double layers of GHAT3 which was proposed and analyzed. We have derived the solution of a two-body problem for an electron and a hole for the model Hamiltonian representing double layer GHAT3. We predict the formation of two types of dipolar excitons, characterized by different binding energies and effective masses, in the double layer of GHAT3.  We have calculated the binding energy, effective mass, spectrum of collective excitations, superfluid density and the mean-field critical temperature of the phase transition to the superfluid state for the two-component weakly interacting Bose gas of A and B dipolar excitons in double layer GHAT3. We have demonstrated that at fixed exciton density, the mean-field critical temperature for superfluidity of dipolar excitons is decreased as a function of the gap $\Delta$. Our results show that $T_{c}$ is increased as a function of the density $n$ and is decreased as a function of the gap $\Delta$ and the interlayer separation $D$.

\medskip
\par
The occupancy of the superfluid state at $T<T_{c}$ can result in the existence of persistent dissipationless superconducting oppositely directed electric currents in each GHAT3 layer, forming a double layer. According to the presented results of our calculations, while the external weak magnetic field, responsible for the formation of the gap $\Delta$ in the double layer of $\alpha-T_{3}$ increases the exciton binding energy,  the mean-field transition temperature to the superfluid phase is increased as the weak magnetic field and $\Delta$ are decreased. Therefore, the dipolar exciton system in a double-layer of GHAT3 can be applied to  engineer a switch, where transport properties of dipolar excitons can be tuned by an external weak magnetic field, forming the gap $\Delta$. Varying a weak magnetic field may lead to a phase transition between the superfluid and normal phase, which sufficiently changes the transport properties of dipolar excitons.

\section*{Acknowledgement(s)}
G.G. would like to acknowledge the support from the Air Force Research Laboratory (AFRL) through Grant No. FA9453-21-1-0046

\end{document}